# Brain inspired neuronal silencing mechanism to enable reliable sequence identification

Shiri Hodassman[1†], Yuval Meir[1†], Karin Kisos[1], Itamar Ben-Noam[1], Yael Tugendhaft[1], Amir Goldental[1], Roni Vardi[2†] & Ido Kanter[1,2*]

[1]Department of Physics, Bar-Ilan University; Ramat-Gan, 52900, Israel.

[2]Gonda Interdisciplinary Brain Research Center, Bar-Ilan University; Ramat-Gan, 52900, Israel.

*Corresponding author. Email: ido.kanter@biu.ac.il

†These authors contributed equally to this work.

**Real-time sequence identification is a core use-case of artificial neural networks (ANNs), ranging from recognizing temporal events to identifying verification codes. Existing methods apply recurrent neural networks, which suffer from training difficulties; however, performing this function without feedback loops remains a challenge. Here, we present an experimental neuronal long-term plasticity mechanism for high-precision feedforward sequence identification networks (ID-nets) without feedback loops, wherein input objects have a given order and timing. This mechanism temporarily silences neurons following their recent spiking activity. Therefore, transitory objects act on different dynamically created feedforward sub-networks. ID-nets are demonstrated to reliably identify 10 handwritten digit sequences, and are generalized to deep convolutional ANNs with continuous activation nodes trained on image sequences. Counterintuitively, their classification performance, even with a limited number of training examples, is high for sequences but low for individual objects. ID-nets are also implemented for writer-dependent recognition, and suggested as a cryptographic tool for encrypted authentication. The presented mechanism opens new horizons for advanced ANN algorithms.**

# Introduction

The activity of neurons as computational elements of the brain depends strongly on their spiking activity history, leading to dynamics with a long-term memory effect[1-4]. This type of neuronal plasticity differs from current approaches in modern machine learning (ML), such as the stochastic dropout technique[5]. In this work, we present new experimental in-vitro results controlling dynamics of the brain, and utilize them to expand supervised ML tasks[6] to include sequence identification, where the input objects have a given order and timing.

The objective of a sequence identification task is to decide whether an embedded sequence is presented as a temporal order input. This is a common task in many aspects of human cognition, including associative memory, the recognition of temporal events, and the identification of handwritten sequences of digits, such as telephone numbers or verification codes. Producing a reliable answer for such a decision problem requires accurate classification of all sequence objects. The realization of a sequence identification task with neural networks usually requires feedback loops[7,8], which suffer from training difficulties and long convergence times. It has become increasingly difficult to use feedforward neural networks for sequence identification because their achievable success rates (SRs) typically range between 0.9 for complex classification tasks[9] and limited datasets[10] to above 0.99 for relatively simple tasks[11]. Because the SR of each object is a finite distance from one, the probability of accurately identifying the entire sequence falls exponentially with the number of objects in the sequence. An additional puzzle is that brain dynamics without the precise and reliable implementation of advanced deep learning techniques[12,13] are expected to achieve significantly lower SRs; nevertheless, they have demonstrated the ability to identify many sequences with high fidelity. Hence, the formation of reliable identification-networks (ID-nets) requires the development of a new type of brain-inspired ML mechanism.

In this work, first, we demonstrate a new brain-inspired mechanism where nodes of spiking feedforward neural networks (SFNNs)[14,15] are temporally silenced following their recent activity. Consequently, transitory objects act on different feedforward sub-networks which are dynamically created by the activity of previous objects. The quantitative results of the ID-nets are then presented for the sequence identification of 10 handwritten digits taken from the Modified National Institute of Standards and Technology (MNIST) database[6,10,16]. Then, we present the preliminary results on the utilization of ID-nets for writer-dependent sequence recognition and their possible application to cryptography[17]. We conclude the theoretical discussion by generalizing the presented results to the same artificial neural network (ANN) architectures, but with continuous activation nodes and training sequences obtained from CIFAR-10[18] on LeNet-5[19], representing a deep convolutional neural network. Finally, the experimental results using neuronal cultures are presented to support the mechanism of temporally silenced nodes. We conclude this paper by summarizing the main results and comparing the presented mechanism of brain-inspired ID-nets with the long-short-term-memory (LSTM)[20,21] approach.

## Results

**Brain-inspired ID-nets: Temporal sequences.** A temporal object order, i.e., a sequence, comprising 10 consecutive handwritten digits taken from the MNIST database is separated using time-lags of $\Delta t$ (Fig. 1a, Methods). Each input digit to the SFNN is represented via $d$ Boolean frames, where the pixels of each frame are probabilistically spiking (white) or non-spiking (black) relative to their gray-level (Fig. 1a). This representation is justified because the time-average of a sufficient number of $d$ Boolean frames is extremely similar to the original digit (Fig. 1a).

A temporal sequence (Fig. 1a) is trained using a backpropagation technique on a feedforward network comprising 784 input units, which represent the 28 × 28 pixels of the MNIST digit, 200 hidden units, and 10 output units, which represent possible labels (Fig. 1b). For a given input digit, the output is selected as the output label with maximal fires in the last $d_1$ frames (Methods). Clearly, the trained network cannot serve as an ID-net because each digit is predicted independently. In addition, with a limited number of examples per label, e.g., 1000, the SR is only ~0.9 (Supplementary Fig. 1), and the probability of accurately identifying all 10 digits thus decreases to $0.9^{10} \sim 0.35$.

**Brain-inspired ID-nets: Silencing mechanism.** The possibility of reconstructing this network as an ID-net emerged from the experimentally observed long-term memory of neurons after they were stimulated at high frequency (see experimental results below), which led to the following dynamical scheme. The response probability, $\frac{q}{d_1}$, of each network node when presented with a given digit depends on the number of its spikes $q$ during the last $d_1$ frames of the previous digit (Fig. 1c). This silencing mechanism is applicable for both input and hidden units, and it is artificially attributed to the first digit to exclude boundary condition effects (Methods). Consequently, each digit is trained on a different sub-SFNN, where the response probability of some of the inputs and hidden units is dimmed based on their past activity (Fig. 1d). The weights that connect the silenced nodes are also temporarily excluded in the training process. In principle, this silencing mechanism has long-range effects, similar to the Markovian processes[22,23]. The activation of nodes for a given digit depends on their activity for the previous digit, which in turn depends on the activity for the penultimate digit; the importance of the temporal order is thus evident. If the order of the test sequence differs from the trained sequence, the active sub-SFNN of some test digits will differ significantly from the trained sub-SFNN (Fig. 1d). When the sequence orders differ, the SRs are expected to decrease following the decrease in the ID-net signal-to-noise-ratio. The signal is diminished because some of the trained weights have been silenced, while the noise is enhanced as some weights that were silenced in the training process have become active.

**Performance of the ID-net.** The performance of an ID-net is examined quantitatively to distinguish between the embedded sequence and following types of imperfections (Fig. 1e). The first imperfection is a much faster presentation of the same sequence with shorter time-lags between consecutive digits $\Delta t_s < \Delta t$, resulting in additional silenced nodes that spiked for the previously presented digits. The second imperfection is a much slower presentation of the same sequence with longer time-lags $\Delta t_L > \Delta t$, such that the effect of the silenced nodes vanished; i.e., all nodes are active. The third imperfection comprises a wrong sequence, where the timings of the two digits have been exchanged.

Training the ID-net (Fig. 1b) using 1000 sequences, wherein each comprises the same order of 10 unique MNIST digits, results in an average SR of ~0.91 per digit. The minimal SR among the 10 digits varies in the range $[0.79, 0.88]$, where the most probable value is near the middle of this range (Fig. 2a, blue histogram). The minimal SR for the three types of imperfect input sequences, i.e., fast, slow, and wrong (Fig. 1e), is summarized by the orange histograms in Fig. 2a. We assume that the ID-net knows the trained order of the digits, and that the SRs are evaluated with respect to the expected order of the digits (Methods). The results clearly indicate a significant gap between the blue and orange histograms for slow and fast imperfections, such that a threshold of 0.6 can be used to identify clusters of such imperfect sequences. For wrong sequences with only two swapped digits, there is a small overlap between the orange and blue histograms (Fig. 2a); however, a threshold of 0.79 accurately predicts whether the order of the input sequence is correct with a probability of ~0.998. The threshold is defined as the maximal value such that the blue histogram, generated from all examined samples, is above it (True positive = 1). Hence, the SR indicates the fraction of the orange histogram obtained from all samples that is below threshold. This definition of SR corresponds to the true negative rate (specificity measurement).

In cases where the ID-net does not know the trained sequence, the SR is calculated for each digit following the most frequently predicted labels. The orange histogram is calculated following the minimal SR value among the 10 digits; consequently, the orange histogram exhibits higher SR values (Fig. 2b). Nevertheless, a significant gap remains for imperfect faster or slower sequences (Fig. 2b); meanwhile, for similar wrong sequences, a threshold of 0.79 accurately predicts whether the order of the input sequence is correct or wrong with a probability of ~0.994 (Fig. 2b).

The ID-net is capable of embedding and recognizing more than one sequence, while remaining robust in terms of identifying imperfect sequences. For the case with two embedded similar sequences (Fig. 2c), there is a slight broadening and shift in the orange histograms, but the gap is maintained for the fast and slow sequences. For wrong sequences a threshold of 0.76 predicts with probability ~0.992 whether the input sequence is one of the correct sequences or not (Fig. 2c), as well as with a probability of ~0.991 when the ID-net does not know the order of trained sequences (Fig. 2d). When increasing the number of embedded sequences, the gap between the orange and blue histograms decreases, and their overlap increases (Supplementary Fig. 2). Note that, for all cases examined here (Fig. 2), a fitted threshold for each trained sequence resulted in an enhanced gap and better identification (Supplementary Fig. 3), because the right tails of the orange and blue histograms were correlated.

The results are shown for sequences of 10 digits and for an architecture with 200 hidden units only (Fig. 1b). Nevertheless, a similar performance of the ID-net was obtained in our simulations for sequences with numbers of digits in the range [5, 12]. The improvement in performance for more than 200 hidden units was found to be negligible, and was only slightly affected in the range [50, 200].

**ID-net for writer-dependent recognition.** Finally, the performance of the ID-net is demonstrated for writer-dependent recognition, which enables a given sequence with specific handwriting to be distinguished from its three associated imperfections and from the same sequence in different handwriting. For simplicity, we classify the handwriting based on the average absolute distance between pixels. Specifically, the maximal absolute difference between the pixel values with a gray-level greater than 100 is 20 for a given handwriting, whereas the handwritings of other individuals comprise other MNIST digits. Indeed, the absolute difference between gray-level pixels greater than 100 of two MNIST examples with the same label is much greater than 20 (Supplementary Fig. 4).

The ID-net (Fig. 1b) is trained on one individual's handwriting using synthetic sequences generated by adding integer noise in the range [-20, 20] to pixels with a gray-level greater than 100. A decision based on the SRs is not applicable in this case because only one sequence is presented with 10 output probabilities for each digit. Hence, for each predicted test digit, the gap, $\Delta$, between the highest and second highest firing output nodes is normalized using $d_1$ to the range $[0, 1]$. Its minimal value for the 10 digits of the sequence $\Delta_{min}$ is chosen for the decision criterion. When the correct sequence is presented, the highest normalized gap $\Delta$ for each digit is expected to be close to one. For a sequence with one of the imperfections or with different handwriting, the two opposite trends are expected in some of the digits, including a decrease in the highest firing output node with a simultaneous increase in the firing of the next highest output node, resulting in a substantial decrease in $\Delta_{\min}$. The minimal gap $\Delta_{min}$ can indeed identify a writer-dependent sequence with a probability approaching 1 (Fig. 3), and it is recommended as a cryptographic authentication ingredient that is robust to noise. The sequence itself is not known to the verification system, which only receives the weights and dynamics of the ID-net for verification. The client must insert the 10 digits in the correct order and speed using correct handwriting, all of which are robust to some level of noise, unlike methods relying on number theory[24]. An opponent is expected to fail even with partial knowledge of the individual handwriting, speed, and order of the sequence (Fig. 3). This capability of ID-net can be extended to identify several writer-dependent sequences (Supplementary Fig. 5). Nevertheless, many open questions remain regarding the level of security of such potential cryptographic applications, which require further research.

**Generalization of ID-nets to ANNs.** The brain-inspired mechanism, where nodes of spiking feedforward neural networks (SFNNs) are temporally silenced following their recent activity, is generalized to standard ANNs with continuous output nodes (Fig. 4a) and to deep convolutional neural networks trained on the CIFAR-10[18] database (Fig. 4b). The silencing probability of each SFNN node when presented with a given object is equal to the fraction of its spikes during the frames of the previous object. Similarly, the silencing probability of an ANN node is determined by its activation value in the previous object. The fully connected architecture comprising a single hidden layer (Fig. 1b) with a sigmoid activation function for the nodes serves as the ID-net using the following dynamical scheme. The silencing probability of each network node when presented with a given MNIST digit is equal to its output [0, 1] in the previous digit. This silencing mechanism is applicable for both the input and hidden units, and it is artificially attributed to the first digit to exclude boundary condition effects (Methods). We train this ANN on a dataset similar to the one in Fig. 2a, and the results for the minimal SR are indicated by the blue and orange histograms for the three types of imperfect test sequences: fast/slow/wrong (Fig. 4a). Results clearly indicate a significant gap between the blue and orange histograms for slow and fast imperfections and a small overlap for the wrong imperfection (Fig. 4a). However, a threshold of 0.9 accurately predicts whether the order of the input sequence is correct with a probability of ~0.994. The application of the ID-net for writer-dependent recognition (Fig. 3) is exemplified for the LeNet-5 architecture with a ReLU (Rectified Linear Unit) activation function. It is trained on one individual's objects, comprising a sequence of 10 different CIFAR-10 objects with 50 additional similar synthetic sequences, where the response probability of each network node in the fully connected layers (layers C5 and F6) depends on its output in the previous digit (Methods). The vanishing overlap between the blue and orange histograms for the four imperfections, fast, slow, wrong, and objects of different individuals (Fig. 4b, similar to Fig. 3), clearly indicates the sensitivity of such convolutional neural networks to writer-dependent recognition. This capability of the ID-net using convolutional ANNs can be extended to identify several writer-dependent sequences (not shown). The generalization of the brain-inspired mechanism from SFNNs to ANNs enhances the connection between ML for sequence identification and advanced cryptographic protocols that are robust to some level of noise.

**Experimental observation of silenced neurons in-vitro.** The inspiration for silencing a node following its recent relatively high-frequency spiking activity originates from our in-vitro experiments on synaptic blocked neuronal cultures[25] (Methods). The experimental setup is capable of repeatedly stimulating a neuron extracellularly via one of its dendrites, while recording the responses of the neuron intracellularly[26] using the patch-clamp technique (Fig. 5a).

Each neuron responds reliably when stimulated above threshold below its critical frequency[27], which varies considerably among neurons. When a neuron is stimulated significantly above its critical firing frequency, its responses can comprise alternating bursts[28] of full responsiveness and silencing periods (Fig. 5b). A burst of full responsiveness can represent repetitive spiking activity in response to consecutive frames of a digit (Fig. 1a), followed by a silencing period (Fig. 1c). This is the underlying mechanism of the present ID-net, where transitory objects act on different dynamically created feedforward sub-networks (Fig. 2d). The silencing periods typically comprise tens of stimulations without evoked spikes, and their maximal durations can reach an order of seconds. The relaxation of such neurons to full responsiveness at low frequency sometimes exceeds one minute[29] (Supplementary Fig. 6).

Notably, although neurons can be silenced for a timescale of seconds, the response of the entire network, beyond the silenced sub-network (Fig. 1d), is controlled by neurons that are active during these periods. Furthermore, sub-second silencing periods were observed for neurons that are characterized by high critical firing frequencies[27], where their responsiveness increases for a given stimulation frequency; hence, these neurons can participate in the identification of high-rate sequences.

## Discussion

In this study, we have demonstrated the complex reality of brain dynamics, wherein the functionality of neurons is time- and activity-dependent. It has been demonstrated that this stochastic long-term memory is not a disadvantage representing biological limitations. Instead, its advantage over the memoryless deterministic dynamics of SFNNs without feedback loops for temporal sequence identification has been presented. Counterintuitively, for a given architecture and limited number of training examples, the classification performance for individual objects without silencing nodes is far below unity, but approaches unity for identification of an entire sequence of objects using silencing nodes. This brain-inspired long-term neuronal plasticity was also useful for realizing sequence identification using ANNs without feedback loops and with a limited number of training examples. It represents the brain-inspired SFNNs as a source of novel mechanisms to advance learning in ANNs.

Note that the underlying mechanism of the present artificial ID-net cannot shed light on the way the brain deciphers the time series of objects. Indeed, recent EEG experiments[30] indicate that each visual cognitive task, i.e., a visual object, activates a different area of the brain, which is highly similar during the same task. However, these brain states dynamically evolve into cyclic manifolds that are distinct with different tasks.

Notably, this type of long-term neuronal plasticity differs from existing memoryless approaches in modern ML, such as the stochastic dropout technique[5]. Here, the temporal dropout of nodal activity depends on their recent activity. In addition, the presented brain-inspired mechanism for silencing nodes differs from the conventional MPEG video compression standards[31-33], where differences between bits of highly similar consecutive frames are evaluated. These different bits, either zero or one, contain temporal information, whereas similar bits are neglected. In this study, active nodes repeatedly fire within frames representing the same object, and are only silenced following their similar activity for the previously presented object.

Currently, the order of sequences of objects can be identified using recurrent neural network (RNN) architectures, and specifically, the scheme of LSTM[20,21]. In contrast to standard feedforward neural networks, LSTM[20,21] includes feedback connections. The presented brain-inspired approach indicates that, fundamentally, sequences can be identified without feedback loops. In addition, sequences are identified beyond the order of their objects also following the time-lags between consecutive objects. This feature is typically beyond the current framework of the LSTM architecture; we suggest its inclusion.

It is not straightforward to compare the performance of LSTM and ID-net models, as they do not perform the same task. For instance, in LSTM, one typically assumes that a sub-sequence is identified accurately, and the task is to measure the SR of the next possible element generated by a given graph[34]. However, in our study, we assume a small training dataset with SRs for an individual object far below unity, and estimate the SR for accurately identifying the entire sequence.

One component of the state-of-the-art sequence recognition models is an attention mechanism[35]. These models outperform[36] the SR of the presented ID-nets, for which the motivation is to provide a bridge between the experimental findings of stochastic neuronal features and advanced machine-learning algorithms. We present an experimental neuronal long-term plasticity mechanism for reliable feedforward sequence identification. The architecture is the same as that for the identification of an individual object, but transitory objects act on different dynamically created feedforward sub-networks. Nevertheless, the presented ID-nets are only useful for identifying very limited types of time series. They are not applicable to analyzing, for instance, video datasets (which are one of the main foci of current research) and time series analysis applications using the attention mechanism[36]. In video datasets, each object is composed of time series comprising many frames that constitute a single object, where frames are related to each other following the content of the video. In addition, the exchange of the first part of the video with the second part possibly does not change its classification. By contrast, an object in the presented ID-nets is limited to a single frame, where the same object cannot be repeated several times in the time series; hence, consecutive objects (frames) significantly vary. In addition, the identification of the limited length time series is sensitive to the order of its constituent objects (several frames). Nevertheless, we expect that integrating such a biological mechanism of silencing nodes in RNN architectures and NN with attention components will enhance their performance and time domain (slow or fast) identification capabilities. Hence, we conclude that experimentally exploring brain dynamics can be a source of new ideas for enhancing ML methods.

## Methods

**Methods of the simulations: Generating MNIST inputs.** Each example, $\tilde{X}^i, i = 1, 2, ..., M$, of the trained dataset consists of 784 pixels, $\tilde{X}_p^i, p = 1, 2, ..., 784$, the values of which represent the gray-level and are in the range [0, 255]. The original 784 pixels are divided by the value 255, representing the firing probability of the pixels. Eventually, $d$ Boolean frames are generated for each example by comparing the firing probability of each normalized pixel $p$ in the frame number $k$ $X_p^k$ to a random number taken from a uniform distribution in the range [0, 1], as described in detail in the references[37].

**Definition of a node as LIF neuron.** Each node in the ID-net (Fig. 1b) is represented by a leaky-integrate-and-fire (LIF) neuron (equations (1)-(3) in reference[16]), with the time constant for membrane potential decay $\tau = 20 ms$.

**Definition of a sequence.** Each sequence consists of 10 consecutive handwritten different digits taken from the MNIST database separated by 200 ms time-lags. Each input digit to the SFNN is represented by $d$ Boolean frames separated by 10 ms. For simplicity, the first digit of each sequence is fixed to be 0, and the rest of the sequence is constructed from the digits 1-9 in a given preselected order.

**Silenced input nodes and hidden nodes.** After each digit is presented as an input consisting of $d$ frames to the ID-net, the silenced nodes for the next presented input digit are selected in the following way. We define a silencing profile to be the probability of each input and hidden node to be silenced depending on its fraction of spiking activity, $\frac{q}{d_1}$, during the last $d_1$ frames of the previous digit (Fig. 1c). Following a selected uniform random number in the range [0, 1] for each node only once, a decision is taken for a node to be either silenced or active in the current $d$ frames of a digit. To exclude boundary condition effects, we artificially attributed silenced input and hidden nodes to the first zero digit in the sequence, using an adopted silencing profile of randomly selected digits of another trained sequence.

Note that the used random seed number for each digit differs between presented trained and test sequences. Hence, the selected silenced nodes in the frames of a given digit are different.

**Feed-Forward and back propagation of the ID-net.** Each node in the ID-net is presented by a leaky integrate-and-fire neuron. We train the ID-net using *N* sequences with the same order of 10 different digits. The time-lag between consecutive frames of a digit is $10$ ms and the time-lag between consecutive digits is $200$ ms, such that the duration of a sequence is $10 \cdot d + 1800$ ms. Note that the voltage of the LIF nodes practically vanishes between consecutive digits.

The following two main features characterize the training process of the ID-net:
1. The ID-net is trained using 1000 sequences with 10 epochs. All sequences have the same order of the 10 different digits, but with different randomly selected examples from the MNIST dataset. A back propagation learning step is done at the end of each presented digit, to minimize the cross-entropy cost function:

$$C = -\frac{1}{M}\sum_{m=1}^{M}\sum_{i=1}^{10}[y_{m,i} \cdot log(a_{m,i}) + (1 - y_{m,i}) \cdot log(1 - a_{m,i})] + \frac{\alpha}{2\eta} \cdot \sum_{j} W_j^2$$

   where $y_{m,i}$ is the desired output value (0 or 1) for the $i^{th}$ output node in the $m^{th}$ input example. It is equal one for the desired label and otherwise zero. $a_{m,i}$ stands for the output of the i's output unit, given by the average firing during the last $d_1$ frames of the *m*'s input example, $\eta$ is the learning rate and $\alpha$ is the regularization coefficient. The outer summation in the left term is over all *M* training examples.
   The summation in the right term is over all weights in the network.

2. The feedforward and the backpropagation are very similar to the common method for the SFNN [37] with the following main modifications:
   - Silenced nodes – terms in the backpropagation including either silenced input nodes or silenced hidden nodes vanish.
   - Only the accumulated output of the last $d_1$ frames is used in the backpropagation process.
   - The presented ID-net consists of a fully connected architecture (a generalization to diluted architectures is straightforward).
   - In case of above-threshold accumulated voltage, a node fires and the voltage is reset to zero, and with vanishing refractory period.

The backpropagation method computes the gradient for each weight with respect to the cost function, and the weights and biases are updated according to the references[16].

**Output label.** The output label of the ID-net for each presented digit is selected following the output label with the maximal fires during the last $d_1$ frames representing the digit. In case two output labels have the same maximal number of fires, the output label is set equal to the first node between the two. Similar results were found for the case where the output label is randomly selected between the two.

**Optimization.** The learning rate, $\eta$, and the regularization coefficient, $\alpha$, are selected to maximize the success rate (SR) of each one of the 10 presented digits in the trained sequences. Note that the maximization of the SRs of the 10 digits is found to be very similar to the maximization of the average SR of the entire sequence. The optimization was based on the cross validation method. In the optimization procedure, we first search a rough grid of the adjustable parameters followed by fine-tuning grids with higher resolutions.

**Error prediction of the ID-net.** Each panel in Fig. 2, Fig. 3 and Fig. 4 consists of blue and orange histograms. The threshold is defined as the maximal value such that the blue histogram, generated from all examined samples, is above it (True positive = 1). The true/false negative/positive results of Figs. 2-4 are summarized in the tables below:

| Fig. 2 panels (a-b), one sequence - Threshold 0.79 | | | |
|---|---|---|---|
| **Panel (a)** | fast | slow | wrong |
| True positive | 1 | 1 | 1 |
| False Negative | 0 | 0 | 0 |
| False Positive | 0 | 0 | 0.002 |
| True Negative | 1 | 1 | 0.998 |
| **Panel (b)** | fast | slow | wrong |
| True positive | 1 | 1 | 1 |
| False Negative | 0 | 0 | 0 |
| False Positive | 0 | 0 | 0.006 |
| True Negative | 1 | 1 | 0.994 |

| Fig. 2 panels (c-d), two sequences - Threshold 0.76 | | | |
|---|---|---|---|
| **Panel (c)** | fast | slow | wrong |
| True positive | 1 | 1 | 1 |
| False Negative | 0 | 0 | 0 |
| False Positive | 0 | 0 | 0.008 |
| True Negative | 1 | 1 | 0.992 |
| **Panel (d)** | fast | slow | wrong |
| True positive | 1 | 1 | 1 |
| False Negative | 0 | 0 | 0 |
| False Positive | 0 | 0 | 0.009 |
| True Negative | 1 | 1 | 0.991 |

| Fig. 3 panels (a-d) - Threshold 0.79 | | | | |
|---|---|---|---|---|
| **Panels (a-d)** | fast | slow | wrong | different |
| True positive | 1 | 1 | 1 | 1 |
| False Negative | 0 | 0 | 0 | 0 |
| False Positive | 0 | 0 | 0.0065 | 0 |
| True Negative | 1 | 1 | 0.9935 | 1 |

| Fig. 4 panel (a), - Threshold 0.9 | | | | |
|---|---|---|---|---|
| **Panels (a)** | fast | slow | wrong | |
| True positive | 1 | 1 | 1 | |
| False Negative | 0 | 0 | 0 | |
| False Positive | 0 | 0 | 0.0056 | |
| True Negative | 1 | 1 | 0.9944 | |
| **Fig. 4 panel (b) - Threshold 0.69** | | | | |
| **Panels (a)** | fast | slow | wrong | different |
| True positive | 1 | 1 | 1 | 1 |
| False Negative | 0 | 0 | 0 | 0 |

| False Positive | 0 | 0 | 0 | 0 |
|---|---|---|---|---|
| True Negative | 1 | 1 | 1 | 1 |

**Figures 2: The calculation of the SRs.** The ID-net SRs are calculated based on a minimum of 200 sequences with the same order of the 10-digits as in the training procedure.
We distinguished between the following two scenarios:
  a. The trained network knows the order of digits in the sequence. Hence, the SR for each one of the 10 presented digits is calculated with respect to the expected digit, even if a wrong sequence was presented (Fig. 1e).
  b. The trained ID-net does not know the order of digits in the trained sequences. In this scenario, for each one of the 10 presented digits in the test sequence the SR is taken as the maximal SR among the 10 possible output units, regardless if the predicted digit differs from the embedded digit in the trained sequence.

The test accuracy in Figs. 2a and 2c were calculated following scenario (a), whereas test accuracy in Figs. 2b and 2d were calculated following scenario (b).
For Figs. 2a and 2c the used parameters were: $\alpha = 3.3 \cdot 10^{-5}$ and $\eta = 2.7 \cdot 10^{-4}$.
For Figs. 2c and 2d the used parameters were: $\alpha = 10^{-6}$ and $\eta = 4.7 \cdot 10^{-4}$.
Each histogram consists of 40 different samples of trained sequences, each with different order of the 10 digits (For simplicity, the first digit was always selected to be zero).
For the slow and fast imperfections, the orange and blue histograms in Fig. 2a and 2b being constructed from 40 data points, one for each sample.
For the wrong sequence in Figs. 2a and 2b, each sample was tested on 36 different possible wrong sequences, differing from the order of the trained sequence by swapping the order of two digits, excluding the first zero digit. Hence, an orange histogram constructed from $40 \cdot 36 = 1440$ data points.

**Two embedded sequences, Figs. 2c and 2d.** The order of the two embedded sequences is very similar. They have the same order except for the swapping of one pair of digits.
The same test procedure as mentioned above was used to construct the histograms for Figs. 2c and 2d, but with 42 different sequences (21 possible wrong sequences for each embedded sequence). Hence, each orange histogram for the wrong imperfection being constructed from $40 \cdot 42 = 1680$ data points. Note that each embedded sequence contributes a data point to the blue histogram (80 data points for 40 samples).

For training sequences with relatively low SR, which occur with very low probability, one can enhance the SR using slightly different $\eta$ for the digit with the lowest SR in the sequence.

**Figure 3.** For a trained sequence, randomly selected from MNIST, 50 similar synthetic sequences representing the same handwriting were generated in the following way. Each synthetic sequence is generated by adding an integer flat noise in the range [-20, 20] for pixels with gray-level greater than 100 in the MNIST sequence. The ID-net was trained using 100 epochs.

Note that each embedded sequence contributes 50 data points to the blue and orange histograms in Fig. 3a, 3b and 3d and the blue histogram in Fig. 3c, a data point for each synthetic data.

The orange histogram in Fig. 3c is constructed similar to Fig. 2 for 36 different possible wrong sequences and 5 synthetic sequences for each of the 40 samples.

For each presented digit, the gap between the highest and the next highest firing output units is defined. This gap is normalized by $d_1$ to the range $[0, 1]$. The minimal value of the normalized gap among the 10 digits, $\Delta_{min}$, is chosen for the decision criterion.

Note that, in Fig. 3, the same initial random seed is used for the formation of each sequence, i.e. each frame. Consequently, similar input sequences generate very similar frames.
The parameters used in Fig. 3 were: $\alpha = 8 \cdot 10^{-6}$ and $\eta = 3.7 \cdot 10^{-4}$.
For trained sequences with extremely low $\Delta_{min}$, which occur with very low probability, one can enhance the $\Delta_{min}$ using slightly different $\eta$ for the digit with the lowest $\Delta_{min}$ in the sequence.

**Figure 4: Architecture and initial weights.** The architecture in Fig. 4a is a fully connected ANN that is identical to Fig. 1b (comprising 784 input units, 200 hidden units, and 10 output units). Each unit in the hidden and output layers has an additional input from a bias unit.

The architecture in Fig. 4b is a modified Le-Net5[19] with convolution layers that operate on each channel separately, where the summation over all colors is performed before the first fully connected layer. Similar results were obtained using the Le-Net5 architecture.

The initial conditions of all weights are randomly chosen from a Gaussian distribution with a zero mean and standard deviation of one.

Input: Each example of the MNIST dataset, $\tilde{X}^i, i = 1, 2, ..., M$, of the trained dataset comprises 784 pixels, $\tilde{X}^i_p, p = 1, 2, ..., 784$, and their values represent the gray-level in the range [0, 255]. Input **X** of the example $\tilde{X}$ comprises the original 784 pixels, where the average pixel value is subtracted from each pixel, and the standard deviation is set to one.

$$X^i = \tilde{X}^i - \frac{1}{784} \sum_{p=1}^{784} \tilde{X}^i_p,$$

$$X^i = \frac{X^i}{std(\tilde{X}^i)}.$$

**Feedforward and backpropagation.** Fig. 4a: The feedforward and backpropagation neural networks used are standard procedures using a sigmoid activation function for the hidden and output nodes. A cross entropy error function is also used.

Fig. 4b: ReLU activation functions are assigned to the hidden and output nodes, and the backward propagation comprises a gradient descent method using a cross entropy error function.

The terms in the backpropagation, including silenced input nodes or silenced hidden nodes, vanish.

**Silenced input and hidden nodes.** Fig. 4a: After each digit is presented as an input, the silenced nodes for the next presented input digit are selected via the following method. We select the silencing probability of each input and hidden node following its activation value, while presenting the previous digit. A uniform random number in the range [0, 1] is selected for each node only once. The decision is made for a node to be silenced or active if the selected random number is lesser or greater than its previous activation value, respectively.

Fig. 4b: The silencing procedure is applied to the nodes belonging to the fully connected layers (C5 and F6). The nodes with activation values greater than one are silenced with a probability of one.

To exclude the boundary condition effects, we artificially attribute silenced input and hidden nodes to the first zero digit in the sequence, using an adopted silencing profile of randomly selected digits from another trained sequence.

The calculation of the SRs is similar to the two scenarios discussed for the ID-net. The test accuracy in Fig. 4a is calculated following scenario (a), whereas the test accuracy in Fig. 4b is calculated following scenario (b).

For Fig. 4a, the parameters used are $\alpha = 5 \cdot 10^{-6}$ and $\eta = 7 \cdot 10^{-2}$.
For Fig. 4b, the parameter used is $\eta = 1 \cdot 10^{-2}$.

Each histogram shown comprises 20 different samples of trained sequences, each with a different order of the 10 digits (for simplicity, the first digit is always set to zero).
For the blue histograms in Fig. 4a, the SRs of each of the 20 samples is tested on 36 different sequences with the same order as in the training procedure, yielding a histogram constructed from $20 \cdot 36 = 720$ data points.
For the blue histograms in Fig. 4b, the SRs of each of the 20 samples is tested on 36 synthetic examples, representing the same handwriting, with the same order as in the training procedure, yielding a histogram constructed from $20 \cdot 36 = 720$ data points.

A similar procedure is used to construct the fast and slow orange histograms of Figs. 4a and 4b, as well as the orange histogram of different handwriting in Fig. 4b.
For the wrong sequences in the orange histograms of Figs. 4a and 4b, each sample is tested on 36 different possible wrong sequences, differing from the order of the trained sequence by the order of two digits being swapped, excluding the first zero digit.
For the different handwriting orange histogram in Fig. 4b, each sequence is tested on synthetic examples, representing the same handwriting (similar to the above procedure in Fig. 3d), with the same order as in the training procedure.

**Materials and Experimental methods (Figure 5).** The In-vitro experimental methods are similar to those of our previous studies[26,38].

**Animals**
All procedures were in accordance with the National Institutes of Health Guide for the Care and Use of Laboratory Animals and Bar-Ilan University Guidelines for the Use and Care of Laboratory Animals in Research and were approved and supervised by the Bar-Ilan University Animal Care and Use Committee.
The study reports in-vitro experiments and is in accordance with ARRIVE guidelines.

**Culture preparation**

Cortical neurons were obtained from newborn rats (Sprague-Dawley) within 48 h after birth using mechanical and enzymatic procedures. The cortical tissue was digested enzymatically with 0.05% trypsin solution in phosphate-buffered saline (Dulbecco's PBS) free of calcium and magnesium, and supplemented with 20 mM glucose, at 37°C. Enzyme treatment was terminated using heat-inactivated horse serum, and cells were then mechanically dissociated mostly by trituration. The neurons were plated directly onto substrate-integrated multi-electrode arrays (MEAs) and allowed to develop functionally and structurally mature networks over a time period of 2-4 weeks in-vitro, prior to the experiments. The number of plated neurons in a typical network was in the order of 1,300,000, covering an area of about ~5 cm$^2$. The preparations were bathed in minimal essential medium (MEM-Earle, Earle's Salt Base without L-Glutamine) supplemented with heat-inactivated horse serum (5%), B27 supplement (2%), glutamine (0.5 mM), glucose (20 mM), and gentamicin (10 g/ml), and maintained in an atmosphere of 37°C, 5% $CO_2$ and 95% air in an incubator.

**Synaptic blockers**

Experiments were conducted on cultured cortical neurons that were functionally isolated from their network by a pharmacological block of glutamatergic and GABAergic synapses. For each culture at least 20 µl of a cocktail of synaptic blockers were used, consisting of 10 µM CNQX (6-cyano-7-nitroquinoxaline-2,3-dione), 80 µM APV (DL-2-amino-5-phosphonovaleric acid) and 5 µM Bicuculline methiodide. After this procedure no spontaneous activity was observed both in the MEA and the patch clamp recording. In addition, repeated extracellular stimulations did not provoke the slightest cascades of neuronal responses.

**Stimulation and recording – MEA**

An array of 60 Ti/Au/TiN extracellular electrodes, 30 µm in diameter, and typically spaced 200 µm from each other (Multi-Channel Systems, Reutlingen, Germany) was used. The insulation layer (silicon nitride) was pre-treated with polyethyleneimine (0.01% in 0.1 M Borate buffer solution). A commercial setup (MEA2100-60-headstage, MEA2100-interface board, MCS, Reutlingen, Germany) for recording and analyzing data from 60-electrode MEAs was used, with integrated data acquisition from 60 MEA electrodes and 4 additional analog channels, integrated filter amplifier and 3-channel current or voltage stimulus generator. Each channel was sampled at a frequency of 50k samples/s, thus the recorded action potentials and the changes in the neuronal response latency were measured at a resolution of 20 µs. Mono-phasic square voltage pulses were used, with an amplitude of

-900 mV and a duration of 0.5 ms (Fig. 5b).

**Stimulation and recording – Patch Clamp**

The Electrophysiological recordings were performed in whole cell configuration utilizing a Multiclamp 700B patch clamp amplifier (Molecular Devices, Foster City, CA). The cells were constantly perfused with the slow flow of extracellular solution consisting of (mM): NaCl 140, KCl 3, CaCl2 2, MgCl2 1, HEPES 10 (Sigma-Aldrich Corp. Rehovot, Israel), supplemented with 2 mg/ml glucose (Sigma-Aldrich Corp. Rehovot, Israel), pH 7.3, osmolarity adjusted to 300-305 mOsm. The patch pipettes had resistances of 3–5 MOhm after filling with a solution containing (in mM): KCl 135, HEPES 10, glucose 5, MgATP 2, GTP 0.5 (Sigma-Aldrich Corp. Rehovot, Israel), pH 7.3, osmolarity adjusted to 285-290 mOsm. After obtaining the giga-ohm seal, the membrane was ruptured and the cells were subjected to fast current clamp by injecting an appropriate amount of current in order to adjust the membrane potential to about -70 mV. The changes in the neuronal membrane potential were acquired through a Digidata 1550 analog/digital converter using pClamp 10 electrophysiological software (Molecular Devices, Foster City, CA). The acquisition started upon receiving the TTL trigger from MEA setup. The signals were filtered at 10 kHz and digitized at 50 kHz. The cultures mainly consisted of pyramidal cells as a result of mainly enzymatic and mechanical dissociation. For patch clamp recordings, pyramidal neurons were intentionally selected based on their morphological properties.

**Extracellular electrode selection**

For the extracellular stimulations in the performed experiments an extracellular electrode out of the 60 electrodes was chosen by the following procedure. While recording intracellularly, all 60 extracellular electrodes were stimulated serially at 2 Hz and above-threshold, where each electrode is stimulated several times. The electrodes that evoked well-isolated, well-formed spikes were used in the experiments.

**Data analysis**

Analyses were performed in a Matlab environment (MathWorks, Natwick, MA, USA). The recorded data from the MEA (voltage) was filtered by convolution with a Gaussian that has a standard deviation (STD) of 0.1 ms. Evoked spikes were detected by threshold crossing, typically -20 mV, using a detection window of [0.5, 30] ms following the beginning of an extracellular stimulation.

**Statistical analysis**

Results (Fig. 5b) were confirmed on 10 experiments using different neural cultures.

**Data availability:** Source data are provided with this paper. All datasets utilized in this study were downloaded from public sources, http://yann.lecun.com/exdb/mnist/ and https://www.cs.toronto.edu/~kriz/cifar.html . Correspondence and requests for materials should be addressed to I.K.

**Author contributions:** S.H. and Y.M. contributed equally to the theoretical part of this work and R.V. is the main contributor to the experimental work. S.H. and Y.M. analyzed the data and prepared the figures. K.K. and I.B. analyzed the Data. R.V. conducted the in-vitro experiments and analyzed the data. Y.T. prepared the tissue cultures and helped with the in-vitro experiments. A.G. contributed to conceptualization. I.K. initiated the experimental and the theoretical study and supervised all aspects of the work. All authors commented on the manuscript.

**Acknowledgments**

I.K. acknowledges partial financial support of the Israel Ministry of Science and Technology, via collaboration between Italy and Israel. S.H. acknowledges the support of the Israel Ministry of Science and Technology.

**Competing interests:** The authors declare no competing financial interests.

**Additional information**
**Supplementary information** The online version contains supplementary material available at sequences are embedded in the ID-net. The order of two sequences is identical, except for the swapping of one pair of digits. (**d**) Similar to **b**, except with two embedded sequences. Each histogram in panels **a** and **b** (**c** and **d**) comprising 40 samples (Methods).

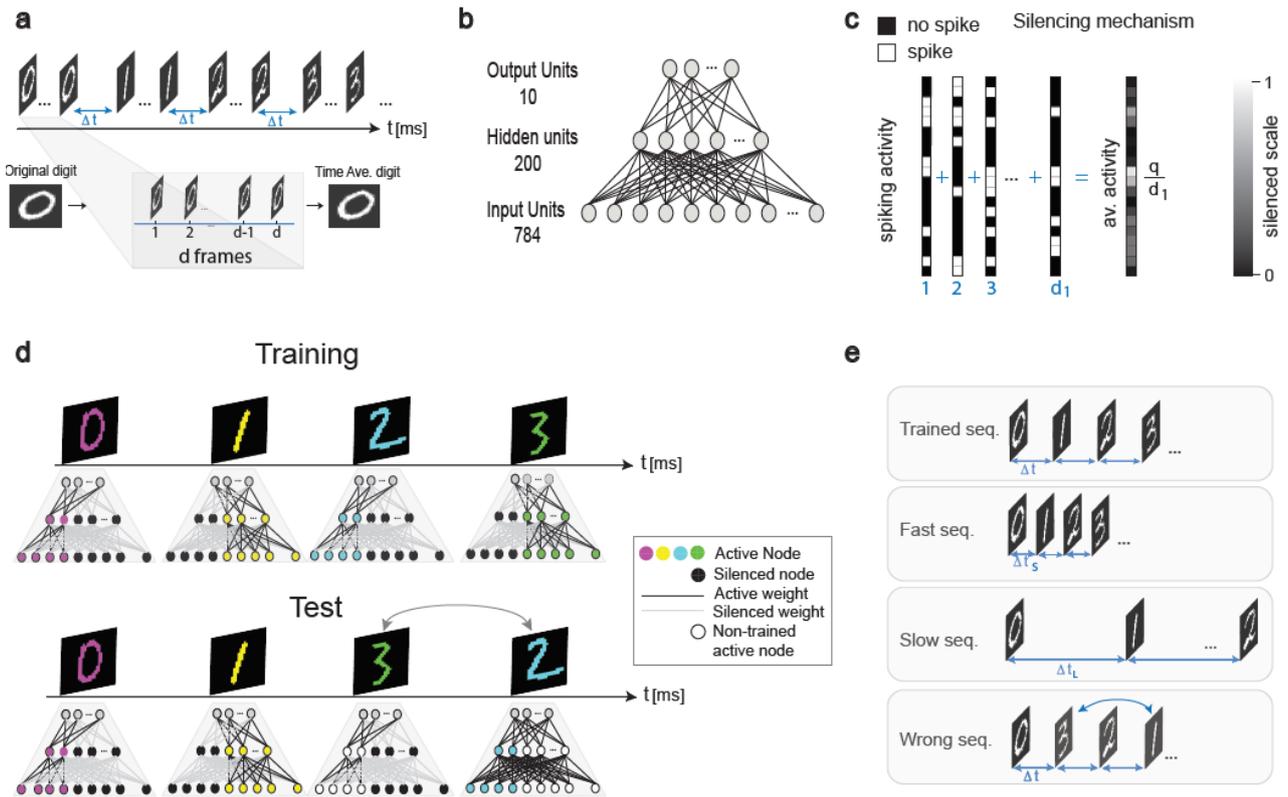

**Figure 1.** Brain-inspired ID-net. (**a**) Scheme of a sequence comprising 10 different MNIST digits separated by $\Delta t$, where each digit in the SFNN is represented by $d$ Boolean frames. (**b**) Fully-connected SFNN architecture comprising 784/200/10 input/hidden/output units, respectively. (**c**) Scheme of the silencing mechanism of each hidden unit following its relative activity, $\frac{q}{d_1}$, in the last $d_1$ frames of the previous digit. The same rule is attributed to the input nodes. (**d**) Training: Scheme where each colored digit is trained on a different sub-SFNN; for simplicity, the nodes are either active (colored following the digit) or silenced (black), and the weights are silenced accordingly. Test: A wrong test sequence where the positions of digits 2 and 3 have been exchanged. The active nodes for digit 3 (white) differ from its trained nodes (green, Training) and corresponding trained weights; the signal deceases and its identification is expected to fail. In addition, white nodes increase noise, because they are active, but non-trained for the presented digit. (**e**) Three possible imperfections—fast/slow/wrong— in test sequences relative to the trained sequence.

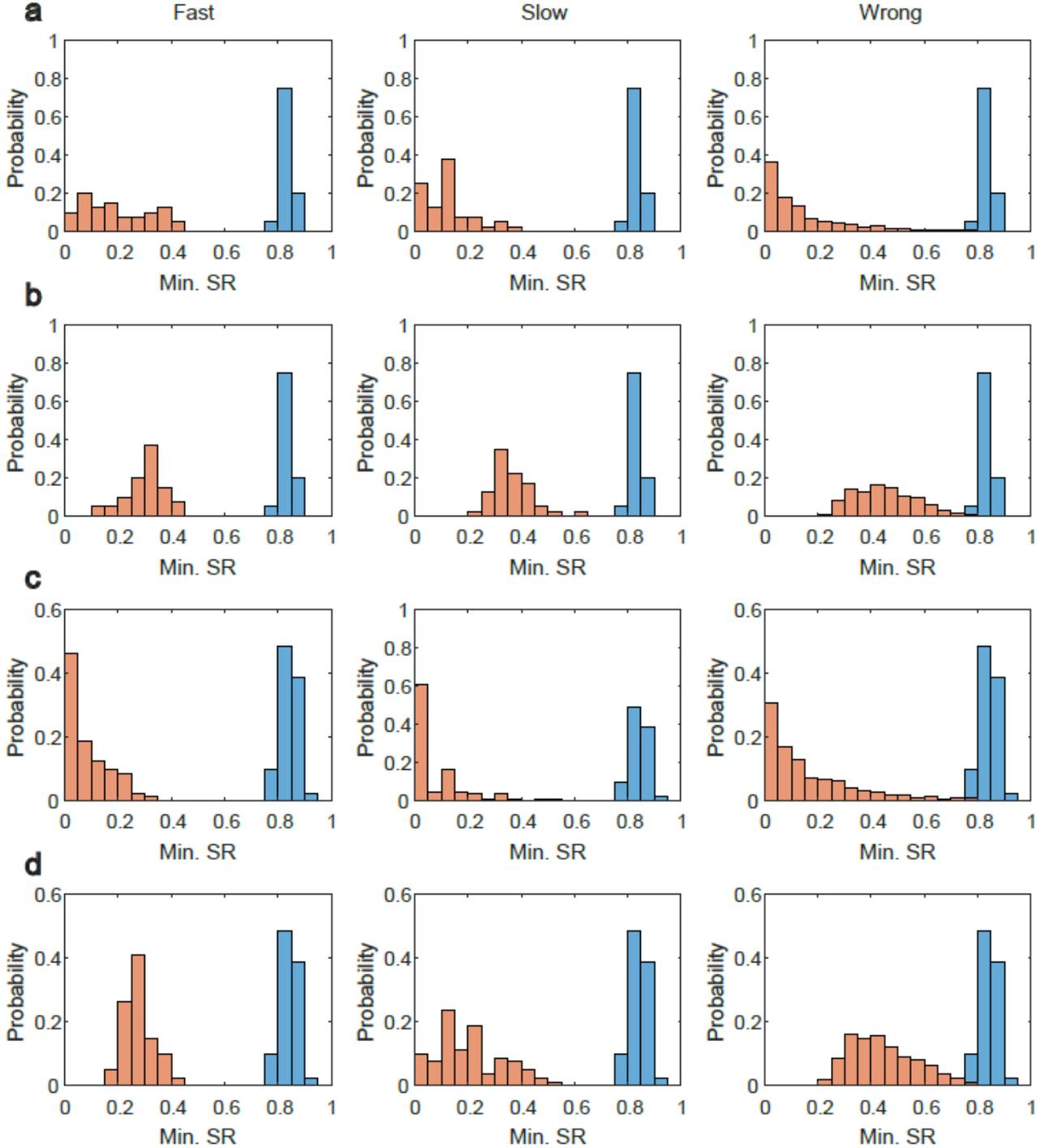

**Figure 2.** Robustness of the ID-net to three types of imperfect sequences. ID-net (Fig. 1b) is trained using 1000 training sequences that each comprises 10 MNIST digits with $d = 30$ and $d_1 = 20$ frames (Fig. 1c), separated by $10\ ms$ with consecutive digits separated by $\Delta t = 200\ ms$ (Methods). (**a**) Blue histogram is constructed from the minimal SR for the individual digits in each trained sequence. Similarly, the orange histograms are constructed from the minimal digit SR for the three types of imperfect test sequences: fast/slow/wrong. (The ID-net knows the order of the trained sequence (Methods). (**b**) Similar to **a**, except that the ID-net does not know the order of the trained sequence. (**c**) Similar to **a**, except that two

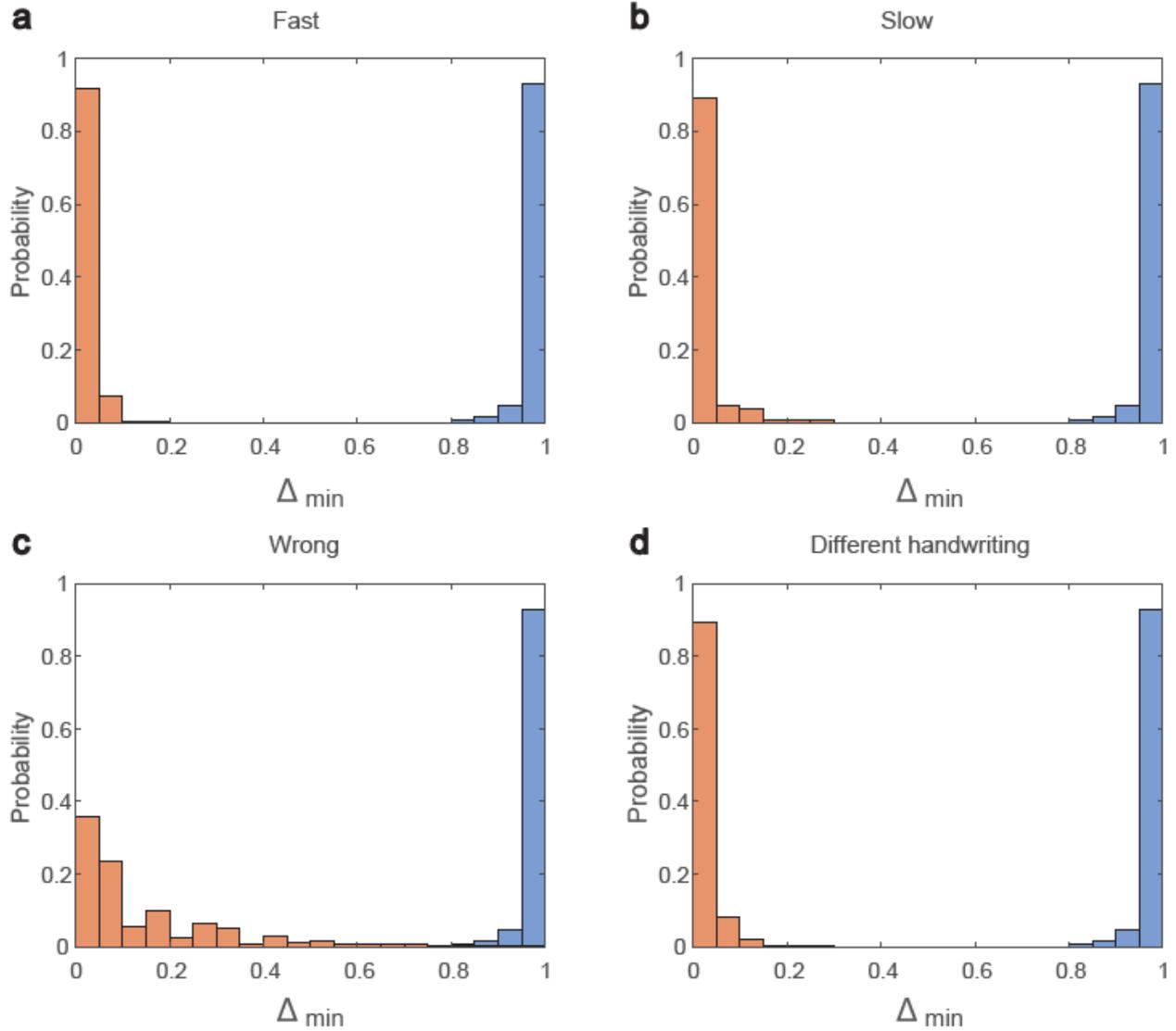

**Figure 3.** ID-net for writer-dependent recognition. ID-net (Fig. 1b) is trained on one individual's handwriting, comprising a sequence of 10 different MNIST digits, with 50 additional similar synthetic sequences generated by adding integer noise in the range [-20, 20] to pixels with a gray-level greater than 100. For each predicted test digit, the gap between the highest and second highest firing output nodes is normalized by $d_1$ to the range [0, 1], and its minimal value among the 10 digits $\Delta_{min}$ is selected (Methods). (**a**) $\Delta_{min}$ for the handwriting of one individual (blue) and for fast sequences (orange). (**b**) Similar to **a** with slow sequences (Fig. 1e). (**c**) Similar to **a** with wrong sequences (Fig. 1e), where a threshold of 0.79 results in a probability of ~0.993 for the accurate classification of input sequences. (**d**) Similar to **a** with the handwriting of different individuals taken from MNIST. A gap is evident between the blue and orange histograms.

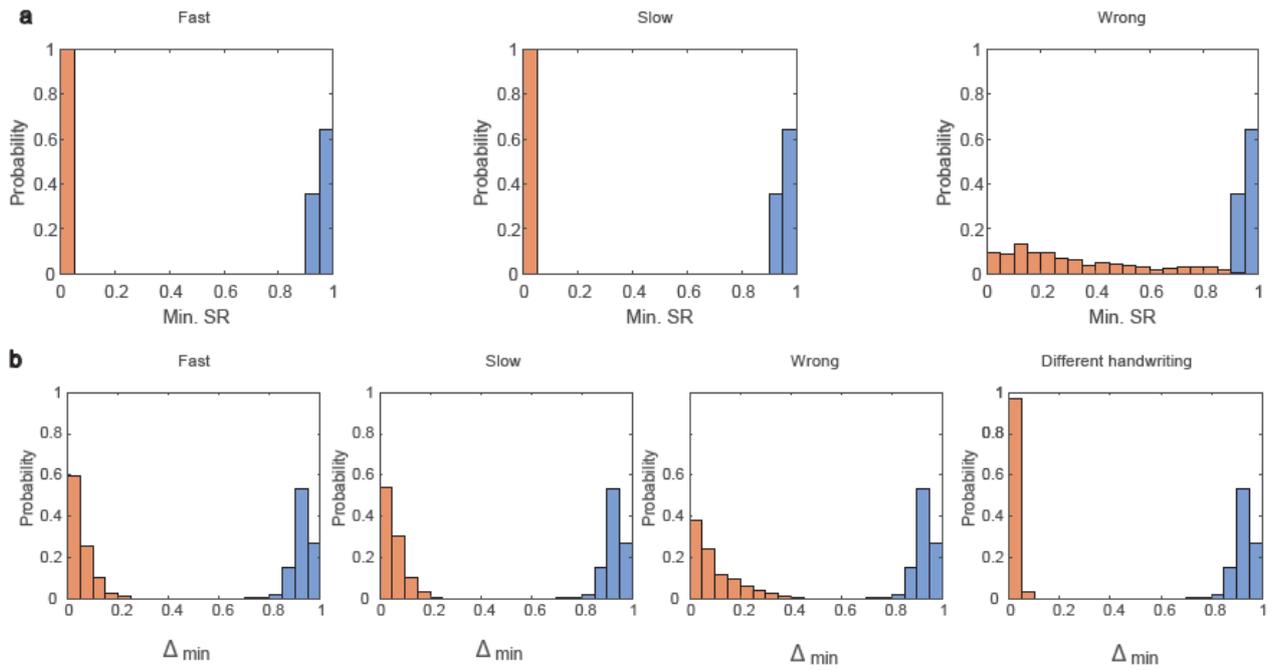

**Figure 4.** Generalization of the brain-inspired SFNNs to ANNs. (**a**) Fully-connected ID-net (Fig. 1b), but with sigmoid activation functions for the units (Methods), which is trained on a dataset that is similar to Fig. 2a (Methods). The blue and orange histograms, averaged over 720 instances, are constructed similarly to Fig. 2a for the three types of imperfect test sequences: fast/slow/wrong. The overlap between the blue and the orange histograms in the wrong panel is 0.56% for threshold 0.9 and zero for the other two panels. (**b**) Similar to Fig. 3, where the LeNet-5 architecture with ReLU activation function is trained on one individual's handwriting, comprising a sequence of 10 different CIFAR-10 objects with 50 additional similar synthetic sequences (Methods). The panels represent the four imperfections: fast, slow, wrong, handwriting of different individuals, similar to Figs. 3a–d. The overlap between the orange and blue histograms vanishes for threshold 0.69, where each histogram was averaged over 720 instances comprising 20 samples (Methods).

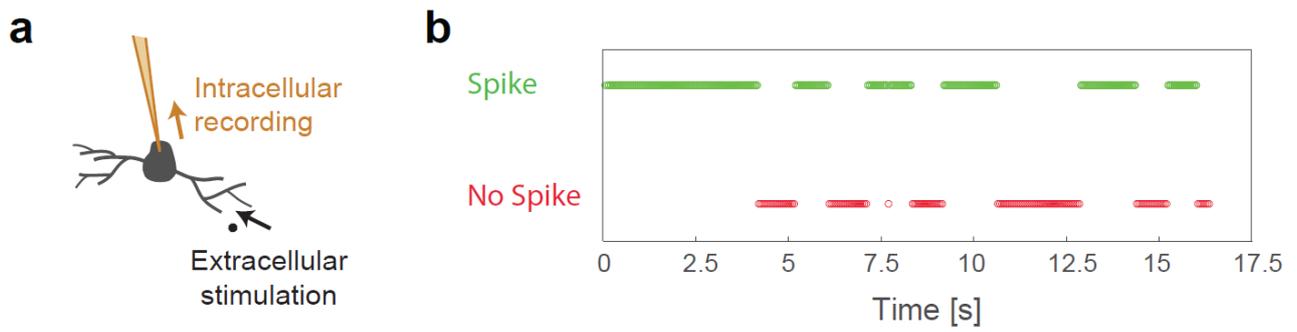

**Figure 5.** Experimental observation of a biological neuron with silencing periods that inspired ID-net. (**a**) Scheme of an in-vitro neuron in a synaptic blocked culture that was stimulated via one of its dendrites and recorded intracellularly. (**b**) Example of recorded neuronal responses, stimulated above-threshold at 20 Hz. The responses comprised time-lags of full responsiveness (segments of green circles) separated by silencing periods (segments of red circles).

# Supplementary Information

## Brain inspired neuronal silencing mechanism to enable reliable sequence identification

Shiri Hodassman[1†], Yuval Meir[1†], Karin Kisos[1], Itamar Ben-Noam[1], Yael Tugendhaft[1], Amir Goldental[1], Roni Vardi[2†] & Ido Kanter[1,2*]

[1]Department of Physics, Bar-Ilan University; Ramat-Gan, 52900, Israel.

[2]Gonda Interdisciplinary Brain Research Center, Bar-Ilan University; Ramat-Gan, 52900, Israel.

*Corresponding author. Email: ido.kanter@biu.ac.il

[†]These authors contributed equally to this work.

**This PDF file includes:**

Figures. S1 to S6

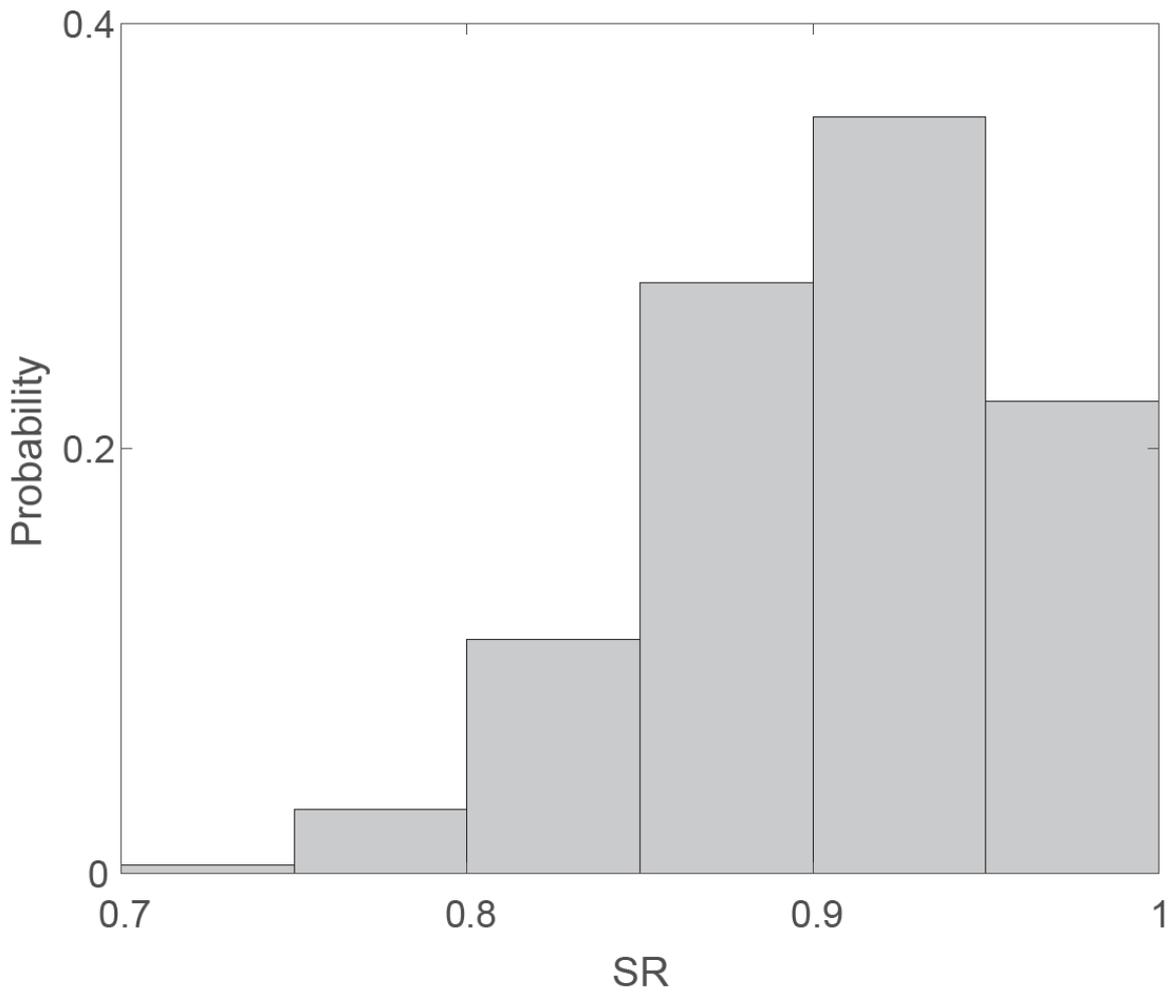

**Supplementary Figure 1.** SRs of the ID-net for one digit without silenced nodes. Histogram of the SRs of the ID-net (Fig. 1b) for individual digits without silenced nodes using 1000 training examples per digit, where $\eta = 7.5 \cdot 10^{-5}$, $\alpha = 1 \cdot 10^{-7}$, and $\mu = 0.9$ (see Methods). The histogram includes the estimated SR for each digit using 1000 test examples, where the measure is repeated 100 times. The averaged SR is ~0.90.

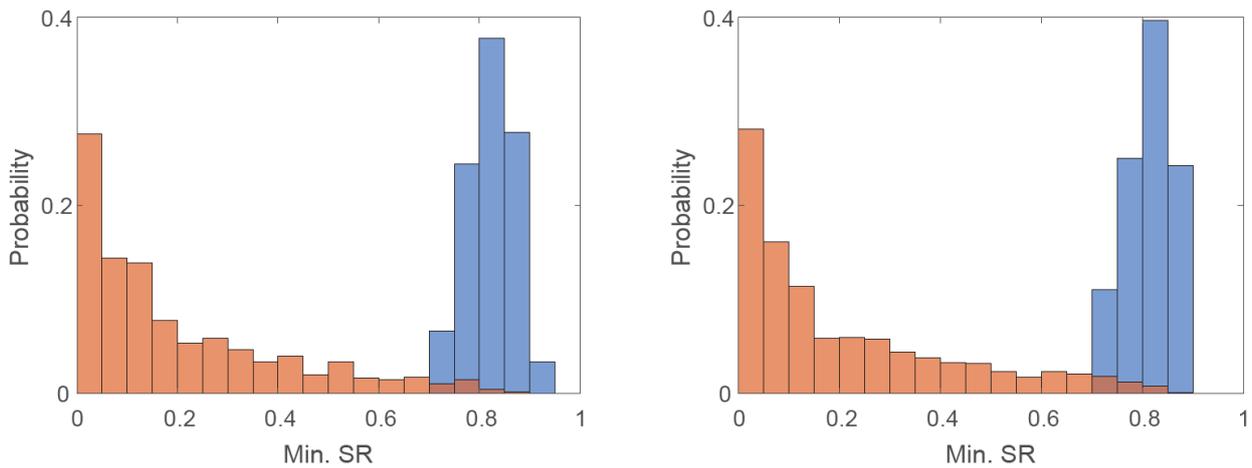

**Supplementary Figure 2.** Robustness of ID-net with multiple trained sequences. The ID-net (Fig. 1b) trained using 1000 training sequences for each embedded sequence, each of which consists of 10 MNIST digits with $d = 30$ and $d_1 = 20$ frames (Fig. 1c) separated by $10\ ms$ with consecutive digits separated by $\Delta t = 200\ ms$. **Left:** Similar to the right panel of Fig. 2a with three trained sequences. There is a small overlap between the orange and the blue histograms, however, a threshold 0.73 predicts correctly with probability ~0.9773 if the order of the input sequence is one of the correct sequences or a wrong one. **Right:** Similar to the left panel of Fig. 1a with four trained sequences. The overlap between the orange and the blue histograms increases, however a threshold of 0.71 predicts correctly with probability ~0.97 if the order of the input sequence is one of the correct sequences or wrong. The mutual order of the multiple embedded sequences is very similar. Each pair of the embedded multiple sequences differs in swapping of one or two pairs of digits only.

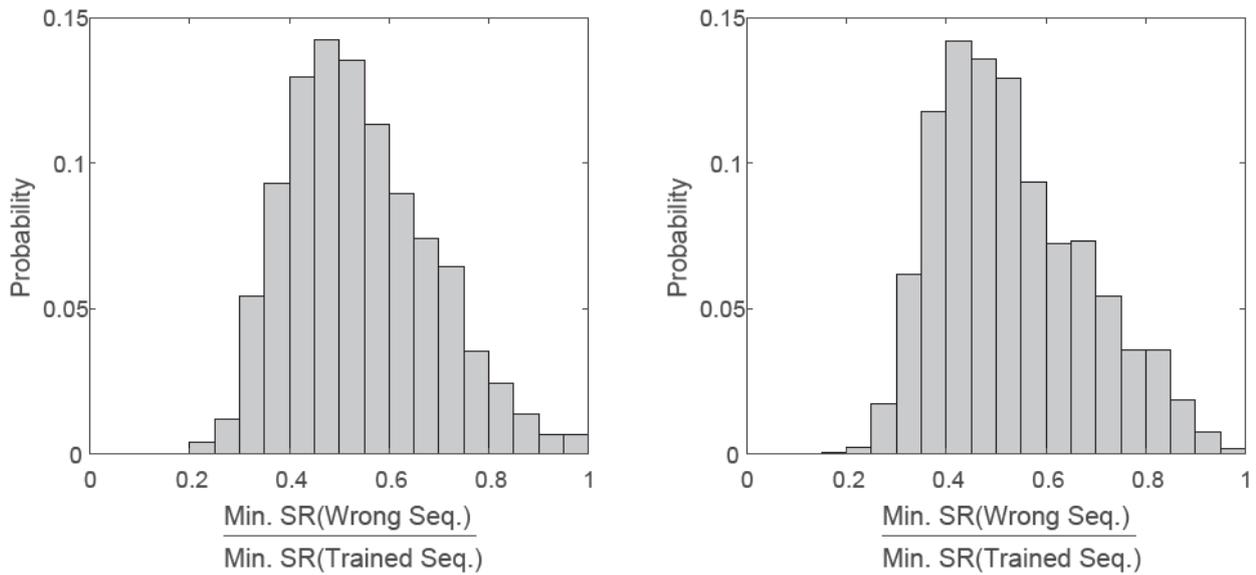

**Supplementary Figure 3.** A fitted threshold for each trained sequence enhances the gap. Histograms for the ratio between the minimal SR for a wrong sequence (Fig. 1e) and for the trained one. **Left:** Histogram for a trained sequence, where the data is taken from the right panel of Fig. 2a. **Right:** Histogram for two trained sequences, where the data is taken from the right panel of Fig. 2b. Since the ratio is less than 1, a fitted threshold for each trained sequence can be used to better distinguish between the embedded sequences and sequences with the one of the three imperfections (Fig. 1e).

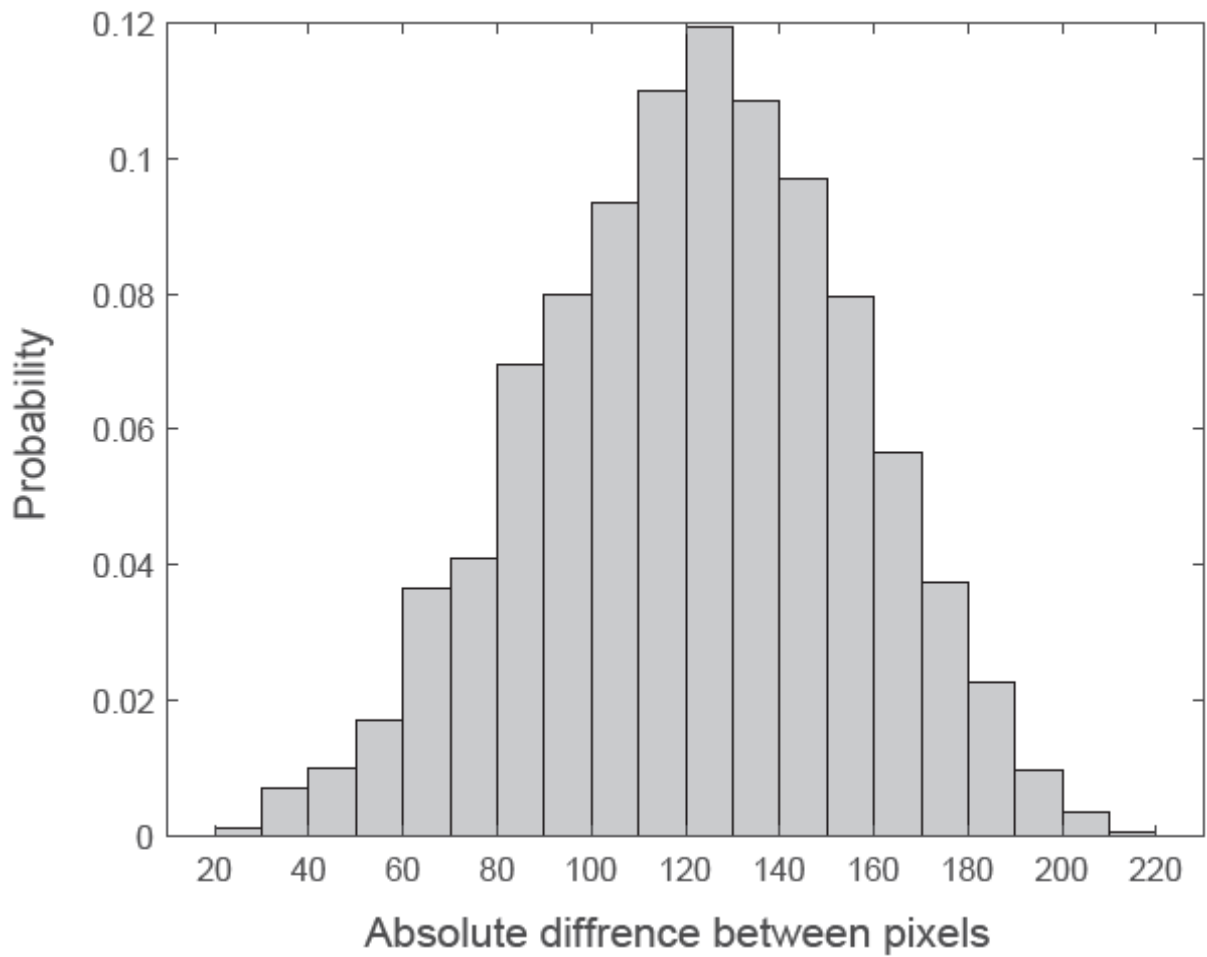

**Supplementary Figure 4.** Histogram of the absolute difference between gray-level pixels greater than 100 of two MNIST examples with the same label. The histogram is constructed from 100 randomly selected pairs for each one of the 10 labels. For each pair, the differences between pixels with gray-level greater than 100 in the first example and the corresponding pixels in the second examples within the pair are calculated. The histogram is built from 1000 pairs, 100 for each label. Note that the minimal difference between two handwritings is greater than 20, where the maximal difference between two digits of the same handwriting is 20, i.e. the average difference is 10 (Fig. 3).

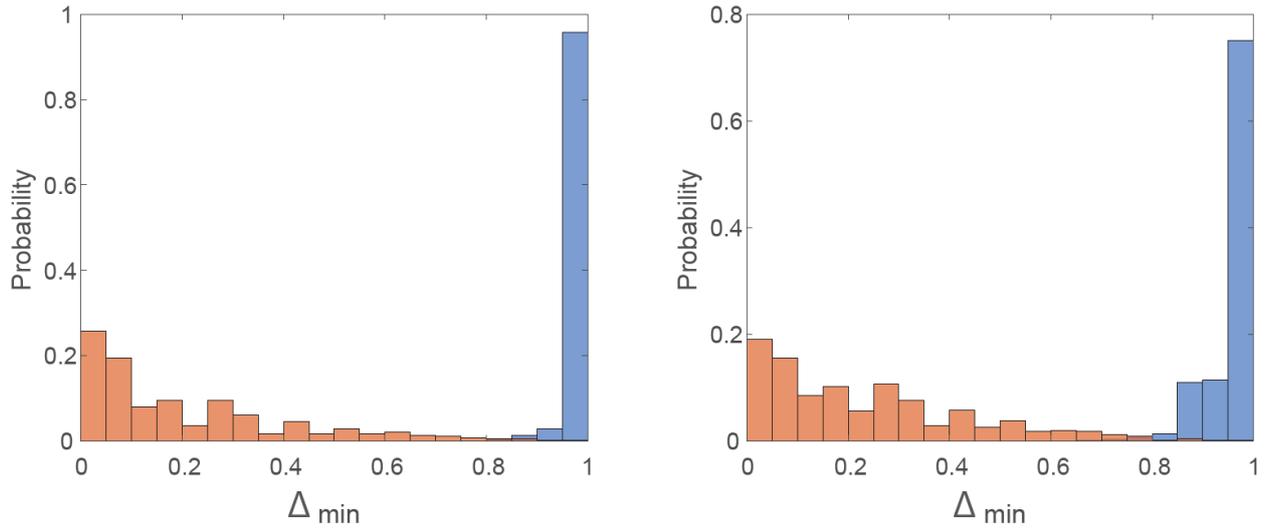

**Supplementary Figure 5.** The capability of ID-net to identify several writer-dependent sequences. The ID-net (Fig. 1b) is trained on several individual's handwritings. The training of each handwriting consists of a sequence of 10 different MNIST digits, with 50 additional similar synthetic sequences generated by adding integer noise in the range [-20, 20] to pixels with a gray-level greater than 100. The 10 digits are ordered differently, such that in the same position along the sequence different digits appear for different embedded sequences. For each predicted test digit, the gap between the highest and next highest firing output nodes is normalized by $d_1$ to the range $[0, 1]$, and its minimal value among the 10 digits, $\Delta_{min}$, is selected. The used parameter are: $\eta = 3.3 \cdot 10^{-3}; \alpha = 1.6 \cdot 10^{-5}$.

**Left panel:** Similar to Fig. 3c with two embedded writer-dependent sequences. A threshold of 0.80 results in a probability of ~0.994 for correct classification of an input as one of the embedded sequences. Note, that we assume a possible opponent has the knowledge of the individual handwriting, the speed of the sequence, as well as the correct position of eight digits among the ten.

**Right panel:** Similar to Fig. 3c with three embedded writer-dependent sequences. A threshold of 0.76 results in accuracy of ~0.985, where accuracy is defined as normalized true-positive plus true-negative.

The left histogram is averaged over $1440 \cdot 2$ instances composed of 40 samples and the right histogram over $1440 \cdot 3$ instances composed of 40 samples.

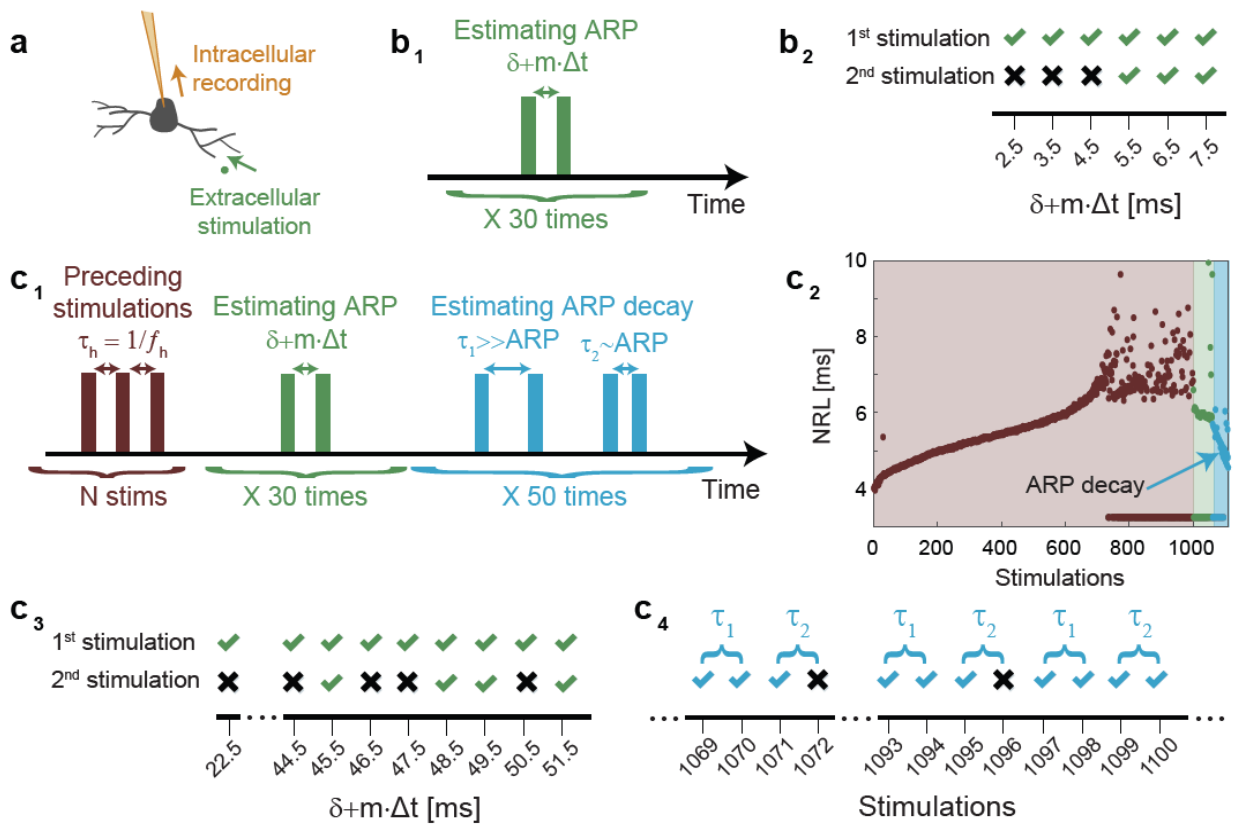

**Supplementary Figure 6.** The relaxation period of the absolute refractory period. (**a**) Scheme of an in-vitro neuron in a synaptic blocked culture that is stimulated via one of its dendrites and recorded intracellularly. (**b$_1$**) Scheme of the experiment measuring the absolute refractory period (ARP) using pairs of stimulation scheduling, with increasing intra-pair time-lags. (**b$_2$**) Neuronal responses for stimulation scheduling at 1 Hz with $\delta$ = 2.5 ms, $\Delta$t = 1 ms in **b$_1$** (V = evoked spike; X = no evoked spike), indicating a 5.5 ms resting ARP. **c** The ARP is a reversible process that after sufficient time returns to its initial value, e.g., 5.5 ms resting ARP (**b$_2$**). The estimation of the decay time of the ARP requires a new type of experiment implementing the following three steps in one stimulation schedule. (**c$_1$**) Scheme of stimulation scheduling that consists of the following three steps: 1000 preceding stimulations (at $f_h$=25 Hz), leading to the intermittent phase (brown); 50 pairs of stimulations (at 5 Hz), with increasing intra time-lags (green) to estimate the stretched ARP ($\delta$ = 22.5 ms, $\Delta$t = 1 ms); and alternating pairs of stimulations (at 0.5 Hz) (blue): one with $\tau_1 = 17\ ms \gg resting\ ARP$ and one with $\tau_2 = 6.5\ ms \sim resting\ ARP$. (**c$_2$**) Neuronal response latency (NRL) and response failures (bottom) for the stimulation scheduling **c$_1$** of neuron **b$_2$**. (**c$_3$**) Stretched ARP = 45.5 ms as measured by the green pairs **c$_1$**. The elongation of the ARP from 5.5 ms to 45.5 ms is an example where the response probability of a neuron is reduced to 50% since the second stimulation within each pair results in response failure. (**c$_4$**)

Responses of the alternating pairs (blue in **c₁**), where responses for the $\tau_2$ pairs start at stimulation 1100, which occurs a minute after the termination of the preceding stimulations (brown in **c₁**); the NRL nearly decays to its resting value (blue in **c₂**).

Results indicate a decay of the stretched ARP back to resting ARP after approximately one minute **c₄**. This relaxation is correlated with the decay of the NRL back to its initial value (**c₂**), as confirmed by examining different neurons (n = 11, mean = 61.5 s, STD = 30.3 s).